# Strain Modulated Electronic and Optical Properties of Laterally Stitched MoSi$_2$N$_4$/XSi$_2$N$_4$ (X=W, Ti) 2D Heterostructures

Ghulam Hussain,[a,*] Mumtaz Manzoor,[b] Muhammad Waqas Iqbal,[b] Imran Muhammad,[c] Asadollah Bafekry,[d,*] Hamid Ullah,[b,*] Carmine Autieri[a]

[a]International Research Centre MagTop, Institute of Physics, Polish Academy of Sciences, Aleja Lotników 32/46, PL-02668 Warsaw, Poland

[b]Department of Physics, Riphah International University, Campus Lahore, Pakistan

[c]School of Materials Science and Engineering, CAPT, Peking University, Beijing 100871, China

[d]Department of Physics, University of Antwerp, Groenenborgerlaan 171, B-2020 Antwerp, Belgium

E-mail: ghussain@ifpan.edu.pl, bafekry.asad@gmail.com, hamid.ullah@riphah.edu.pk

## ABSTRACT

We used first-principles calculations to investigate the laterally stitched monolayered MoSi$_2$N$_4$/XSi$_2$N$_4$ (X=W, Ti) 2D heterostructures. The structural stability of such heterostructures is confirmed by the phonon spectra exhibiting no negative frequencies. From the electronic band structures, the MoSi$_2$N$_4$/WSi$_2$N$_4$-lateral heterostructure (MWLH) shows semiconducting nature with an indirect bandgap of 2.35 eV, while the MoSi$_2$N$_4$/TiSi$_2$N$_4$-lateral heterostructure (MTLH) revealed metallic behavior. Moreover, the effect of biaxial strain on the electronic and optical properties of MWLH is studied, which indicated substantial modifications in their electronic and optical spectra. In particular, an indirect to direct bandgap semiconducting transition can be achieved in MWLH via compressive strain. Besides, the absorbance, transmittance and reflectance spectra can effectively be tuned by means of biaxial strain. Our findings provide insights into the strain engineering of electronic and optical features, which could pave the way for future nano- and optoelectronic applications.

## 1. Introduction

The discovery of graphene,[1-3] has given rise to tremendous research interest in layered 2D materials owing to their excellent physical properties and remarkable potential in device applications.[4-7] Owing to the unique optoelectronic properties, high carrier mobility, and on/off ratio, the 2D transition metal dichalcogenides[8-11] and phosphorene[12-15] are widely explored in the last decade. In order to fully exploit these semiconducting layered structures in the nano- and optoelectronics, the tuning of the bandgap is crucial and plays a vital role in the device application. One of the approaches to modulate the bandgap is to apply a tensile or compressive strain to the 2D layered materials.[16-20] This way, one can tune the physical properties of 2D and quasi 2D materials such that strain-sensitive devices could be realized.[9, 21-30] In 2020, the two 2D layered materials $MoSi_2N_4$ and $WSi_2N_4$ were experimentally synthesized with chemical vapor deposition (CVD) method.[31] The results demonstrated semiconducting characteristics with outstanding mechanical strength and ambient stability. Also, an intrinsic mobility of 270 $cm^2$ $V^{-1}s^{-1}$ for the electrons and of 1200 $cm^2$ $V^{-1}s^{-1}$ for the holes were calculated for $MoSi_2N_4$, which is about four times greater than $MoS_2$.[32] With such remarkable features, $MoSi_2N_4$ and its other group members $MA_2Z_4$ 2D monolayers (where M stands for transition metal Mo, W or Ti, A for Si or Ge and Z represents N, P, or As) are widely studied recently.[33-43] Also, the effect of strain in this class is investigated, highlighting the modification of their electronic and optical properties.[42, 44-48] Nevertheless, exploring the biaxial strain in lateral heterostructures (LH) of this class is essential for future applications, which is missing and not reported to date.

In this paper, we study the structural stability, electronic and optical properties of the $MoSi_2N_4/WSi_2N_4$-lateral heterostructure (MWLH) and $MoSi_2N_4/TiSi_2N_4$-lateral heterostructure

(MTLH) using first-principles calculations. Our results illustrate significant modifications in the electronic band structures with the application of mechanical strain, for instance, the MWLH demonstrates a considerable decrease (increase) in the bandgaps when tensile strain (compressive strain) is applied, and intriguingly, indirect to direct gap semiconductor transition appears via a compressive strain. Moreover, the optical properties are substantially modulated through the application of biaxial strain. The tuning of electronic and optical features in the laterally stitched monolayered heterostructures via strain engineering could offer interesting possibilities in nano- and optoelectronics.

## 2. Methodology

The first-principles calculations were performed via the Vienna ab initio simulation package (VASP)[49, 50] using relativistic density functional theory (DFT). The core and the valence electrons were treated within the projector augmented wave (PAW) method[49] with a cutoff of 550 eV for the plane-wave basis. For relaxation, a mesh of $15 \times 15 \times 1$ k-points was chosen. The periodic images of heterostructures were separated by a vacuum layer of ~20 Å in the perpendicular direction to the plane of monolayered LH. This would avoid spurious interactions among the periodic images. Due to the importance of spin-orbit coupling, maximum calculations were performed with the relativistic effects taken into account. We set the convergence criteria to 0.01 eV/Å and energy $10^{-7}$ eV for our calculations. Also, the dynamical stability of $MoSi_2N_4$/$WSi_2N_4$-LH and $MoSi_2N_4$/$TiSi_2N_4$-LH was investigated through PHONOPY code using a $4\times4\times1$ supercell.[51]

The dielectric function is dependent on the thickness of vacuum, when the two-dimensional (2D) materials are simulated with sufficiently large distances between the periodic images to avoid the interactions between the 2D layers in DFT calculations.[52, 53] This thickness problem is circumvented by characterizing the optical properties of 2D materials using the optical conductivity $\sigma_{2D}(\omega)$. From Maxwell equation, one can express the 3D optical conductivity as;[54] $\sigma_{3D}(\omega) = i[1 - \varepsilon(\omega)]\varepsilon_o\omega$, here $\varepsilon(\omega)$ represents the complex dielectric function $\varepsilon(\omega) = \varepsilon_1(\omega)+i\varepsilon_2(\omega)$, $\varepsilon_o$ denotes the permittivity of vacuum layer, and $\omega$ signifies the frequency of incident photon. The 2D optical conductivity $\sigma_{2D}(\omega)$ is related to $\sigma_{3D}(\omega)$ as;[54, 55] $\sigma_{2D}(\omega) = L\sigma_{3D}(\omega)$, where L characterizes the slab thickness. The normalized absorbance $A(\omega)$, transmittance $T(\omega)$ and reflectance $R(\omega)$, do not depend on the polarization of light for a freestanding 2D layer if normal incidence is considered.[54, 55]

$$A = \frac{Re\ \tilde{\sigma}}{|1 + \tilde{\sigma}/2|^2}$$

$$T = \frac{1}{|1+ \tilde{\sigma}/2|^2} \quad (1)$$

$$R = \left|\frac{\tilde{\sigma}/2}{1 + \tilde{\sigma}/2}\right|^2$$

where $\tilde{\sigma}(\omega) = \sigma_{2D}(\omega)/\varepsilon_o c$ represents the normalized conductivity and c is the speed of light. The above equations are valid for semiconducting 2D materials with the constraint of T+A+R = 1.

### 3. Results and discussions

*Structure and stability*

The laterally stitched monolayered heterostructures ($MoSi_2N_4/WSi_2N_4$ and $MoSi_2N_4/TiSi_2N_4$) crystallize in the hexagonal honeycomb structure. Fig. 1(a) depicts the side and top views of these heterostructures exhibiting the stacking of seven atomic layers of N-Si-N-M-N-Si-N, where M =Mo, W and Ti. The strong bonding of these seven layers of atoms gives rise to a sandwich structure, comprising alternative $MoN_2$ and $WN_2$ ($MoN_2$ and $TiN_2$) layers and two SiN bilayers. We noticed the optimized lattice constants for MWLH and MTLH to be 5.824 and 5.841 Å, respectively. The dynamical stability of MWLH and MTLH is studied through the phonon dispersion calculations as shown in Fig. 1(b), and (c). We employed the finite difference method to calculate the phonon spectra using the Phonopy code[51] with VASP[56] as a calculator. Due to the interface heterostructure, we have a Brillouin zone where the K- and M-points are not all equivalent. We have two inequivalent M-points and two inequivalent K-points like in uniaxially strained graphene.[57] We define these points as M1, M2, K1 and K2. From the phonon band structures along the Brillouin zone's high-symmetry directions Γ-K1-M1-K2-M2- Γ shown in Fig. 1(b) and (c), we can see no imaginary frequencies. From similar results along other directions of the **k**-space, it is established that both the MWLH and MTLH systems are dynamically stable due to the absence of imaginary phonon frequencies in the entire hexagonal Brillouin zone.

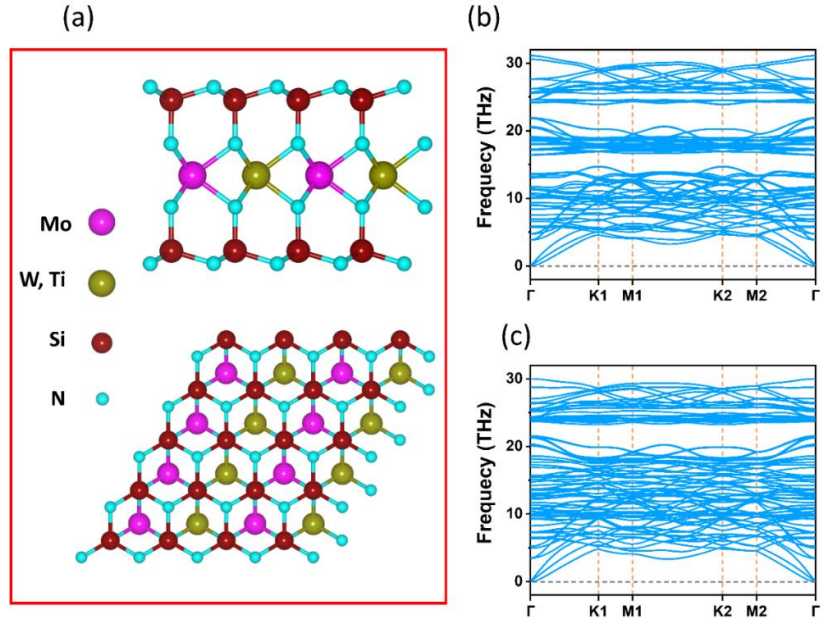

**Figure 1** (a) Side and top views for optimized structure of laterally stitched monolayered heterostructures. (b, c) Phonon spectra of MWLH and MTLH along the path Γ-K1-M1-K2-M2- Γ. The phonon spectra indicate that both MWLH and MTLH are dynamically stable.

*Electronice properties*

We report the electronic bandstructure of the MWLH and MTLH on the complete k-path including M1, M2, K1 and K2, as illustrated in Fig. 2(a) and supplementary Fig. S1. Our results show that the conduction band minimum (CBM) appears at K1 and K2, whereas the valence band maximum (VBM) is at Γ-point, signifying the indirect nature of the bandgap for MWLH. Since we have established that the band structure of the system at M1 and M2 as well as at K1 and K2 is almost equivalent, in the rest of the paper we will focus on the reduced **k**-path that includes just M1 and K1, and that will be renamed as M and K in the rest of the paper. Figure 2(a) manifests that CBM (at K1/K2) and VBM (at Γ-point) in MWLH are almost equally contributed by Mo and W orbitals. The size of the dot shows the amount of contribution from each atom (Mo states are denoted by purple dots, while W states are shown by green dots), the bigger the size of the dot, the more the orbitals of that atom contribute. Likewise, the band structure for the MTLH on the complete k-path (Γ-K1-M1- Γ-K2-M2) is shown in Fig. S1. However, the structure shows metallic behavior, where the conduction and valence bands cross the Fermi-level along Γ-K2 direction. Since the PAW method based on PBE underestimates the electronic bandgap, therefore more accurate method such as HSE06 is used to calculate the bandstructure. In figure 2(b), we compare the electronic bandstructures obtained through the the generalized gradient approximation with PBE form and Heyd-Scuseria-Ernzerhof (HSE06) denoted by black and blue colors, respectively. The bandgaps are approximated to be 1.92 eV for PBE and 2.35 eV based on HSE06 functionals.

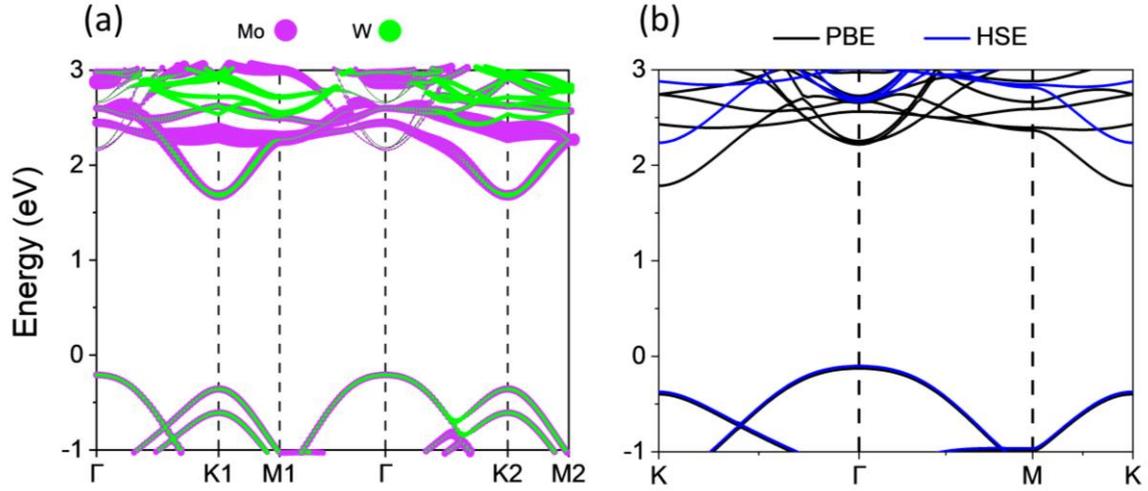

**Figure 2** (a) Projected bandstructure of MWLH along the complete k-path (Γ-K1-M1- Γ-K2-M2), the Mo (purple) and W (green) atoms almost share equal contributions to the VBM and CBM. (b) Electronic bandstructures of MWLH calculated with PBE (black) and HSE06 (blue).

*Strain modulated electronice bandstructures*

To investigate qualitatively, the effect of biaxial strain on the electronic properties of the monolayered MWLH, the electronic band structures are computed along the reduced **k**-path K-Γ-M-K, where K=K1 and M=M1 using PBE potential. We accomplished this by applying successive compressive and tensile strains to the monolayered heterostructures ranging from -5 to 5%, respectively. The variation of electronic band structures is studied in detail and the most important results are demonstrated in Fig. 3 for MWLH. Figure 3(a) illustrates the band structure for strainless MWLH, the valence band maximum (VBM) occurs at 'Γ' point, while the conduction band minimum (CBM) appears at 'K' point of the Brillouin zone suggesting indirect nature of the bandgap. The red, green and purple dots in the bandstructure plots indicate the position of VBM and CBM. We then apply tensile and compressive strains in the range of -5 to 5%, to inspect their influence on the band structure. As expected, the interatomic distances are greatly modified. The bond lengths of Mo-N ($d_{Mo-N}$), W-N ($d_{W-N}$), and also that of Si-N ($d_{Si-N}$) vary considerably with the biaxial strain (Table I). These variations in the bond lengths have a substantial impact on the electronic band structures of the strained systems. Upon applying the tensile strain, we noticed that the electronic bandgap significantly decreases from 1.92 eV (unstrained) to 0.95 eV (5% strained) as the value of strain increased. However, it still preserves the nature of the indirect bandgap up to 10%. This evolution in the bandgap is associated with the shifting of the energy states of the conduction band at the K point of the Brillouin zone. From the analysis of the partial density of states (shown in the supplementary materials, Figure S2(a)), it is evidenced that these states are mainly dominated by the 'd' orbitals of Mo and W. Due to the application of biaxial strain, the bond lengths among the atoms change significantly that causes a different superposition of atomic orbitals, which results in the shift of energy states. A similar trend of variation in the electronic band structure has previously been observed in the layered structure of $MoS_2$, when the distance between the Si–Mo–S sheets and their number were changed.[58] Figures 3 (e)-(g) depict the electronic band structures of MWLHs for biaxial compressive strain. Up to -4% compression, we found a monotonic increase in the bandgap, however, beyond that, the bandgap started to decrease (ranging from 2.14 – 2.4 eV). In addition, the indirect bandgap behavior was retained up to -2% compression, nonetheless, the nature of the bandgap changes from the indirect to direct at approximately -3% compressive strain. During the transition, we noted that VBM shifts from 'Γ' to the 'K' point in the Brillouin zone. The variation of the bandgap and the transition from indirect to direct bandgap semiconductor is caused due to the shifting of energy states near the VBM and CBM at the high-symmetry **k**-point of the Brillouin zone. To determine the type of orbitals that are involved in causing these variations, the partial density of states (PDOS) for unstrained and strained (-3%) is computed as shown in the supplementary materials Fig. S2(a), and (b), which specify the contribution of orbitals from different atoms. For strainless MWLH, we noted that the energy states near the VBM and CBM originated mainly from the Mo-/W-d orbitals with little contribution N-p orbitals. However, at the transition point (-3%), the N-p orbitals of VBM at the 'K' become dominant with respect to the Mo-/W-d orbitals. Figure 2(h) shows the overall modulation of the electronic energy bandgap by the biaxial strain. The black, cyan and orange circles represent the strainless, compression and tensile strains' studies, respectively. The main reason behind the strain-induced band structure engineering is the changes in the interatomic distances via lattice expansion or compression, see Table I.

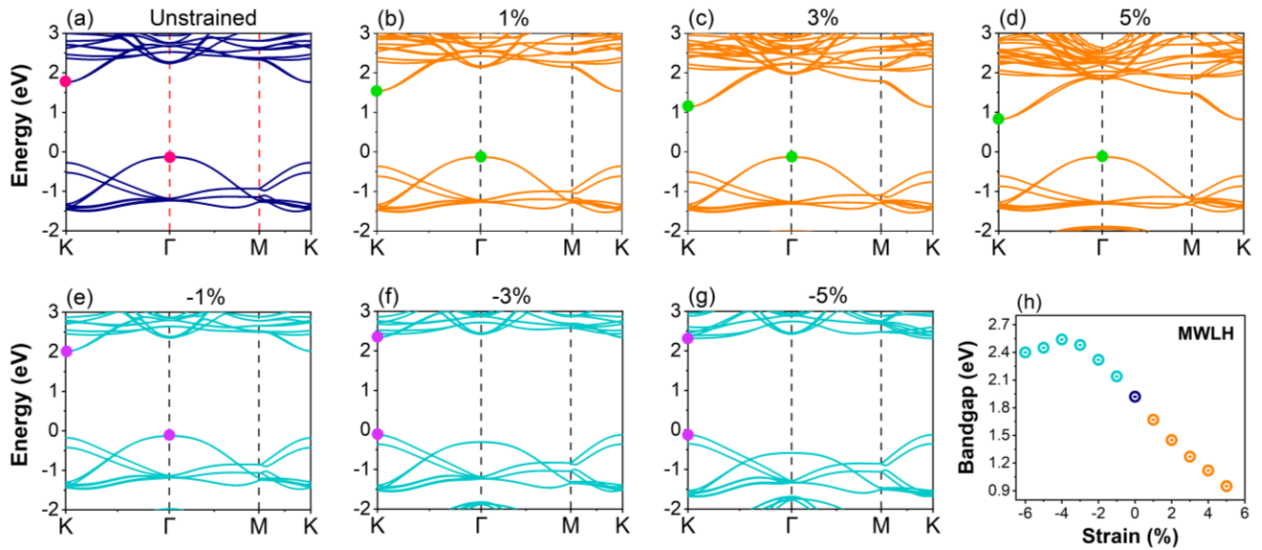

**Figure 3** Electronic band structures of MWLHs along the reduced k-path (K-Γ-M-K), with spin-orbit coupling taken into account. The navy line in (a) demonstrates the bandstructure of unstrained MWLH, while the orange curves in (b), (c) and (d) show the bandstructures for successive tensile biaxial strains. Figures (e)-(g) represent the bandstructures of MWLH against compression. The red, green and purple dots indicate the position of VBM and CBM. (h) The trend of electronic bandgap against the compression and tensile strains, respectively.

10Since $MoSi_2N_4$ materials are known to host gapped states in a pair of valleys, which are found at the corners of the Brillouin zone.[59-61] Owing to the breaking of inversion symmetry in these monolayers, the spin states become separate in energy resulting in spin-valley couplings near the Fermi level and orbital magnetic moments and valley-contrasting Berry curvatures.[62-67] By performing the bandstructure calculations without including spin-orbit coupling (SOC), the states at all the high-symmetry points were observed to be two-fold spin degenerate (as shown in Fig. S3). However, by including SOC, the top of the valence bands at K display spin-splitting of ~125 meV. In contrast, the valence bands occurring at 'M' and 'Γ' points maintain their two-fold spin degeneracy. We noted that the spin states remain separate in energy at 'K', and are not influenced by the biaxial strain indicating the robustness of spin-splitting via SOC. This offers a wide range of potential applications in valleytronics and optoelectronics.[58, 59, 65, 67]

**Table I** Calculated interatomic distances of W-N ($d_{W-N}$), Mo-N ($d_{Mo-N}$), and Si-N ($d_{Si-N}$) for lateral heterostructure of $MoSi_2N_4/WSi_2N_4$ as a function of the applied strain.

| Tensile strain (%) | $d_{W-N}$ (Å) | $d_{Mo-N}$ (Å) | $d_{Si-N}$ (Å) | Compression (%) | $d_{W-N}$ (Å) | $d_{Mo-N}$ (Å) | $d_{Si-N}$ (Å) |
|---|---|---|---|---|---|---|---|
| 0.00 | 2.09 | 2.10 | 1.76 | 0.00 | 2.09 | 2.10 | 1.76 |
| 1.00 | 2.10 | 2.12 | 1.77 | -1.00 | 2.09 | 2.09 | 1.74 |
| 2.00 | 2.11 | 2.13 | 1.78 | -2.00 | 2.08 | 2.08 | 1.73 |
| 3.00 | 2.12 | 2.14 | 1.79 | -3.00 | 2.07 | 2.07 | 1.72 |
| 4.00 | 2.13 | 2.15 | 1.81 | -4.00 | 2.06 | 2.06 | 1.71 |
| 5.00 | 2.14 | 2.16 | 1.82 | -5.00 | 2.05 | 2.05 | 1.69 |

*Strain modulated optical properties*

2D materials are usually known to sustain lattice strains, providing the possibility to effectively tune their properties via strain engineering. In this section, we report the impact of the biaxial strain on the optical properties of MWLH. The application of tensile or compressive strains on the MWLHs subsequently modulates the optical properties. The dielectric function $\varepsilon(\omega)$ for the heterostructure in the supercell geometry having vacuum level on both sides is estimated using random phase approximation (RPA) level. Afterwards, we converted the $\varepsilon(\omega)$ to in-plane conductivity $\sigma_{2D}(\omega)$, and then employed Eq. 1 to estimate the absorbance, transmittance and reflectance.[68] In Fig. 4(a) and (d), the optical absorbance spectra of MWLH for the respective tensile and compressive strains are demonstrated. For the unstrained case, we noticed almost 0% absorbance upto 1.5 eV, however, beyond that the aborption successively increases. A red-shift (blue-shift) in the absorbance of the heterostructure can be observed with increasing the tensile strain (compressive strain), suggesting the sensitivity and fine-tuning of the optical properties of these structures through biaxial strain. The first peak of the absorbance spectrum occurs at ~2.25 eV for strainless MWLH (specified by the black curve in Fig. 4). This can be related to the direct optical transition at 'K' point of the Brillouin zone (see the band structure plot of Fig. 2(a)).[69] The first absorption peak appearing at 2.25 eV for unstrained MWLH, is redshifted to 2.16, 2.01, 1.89, 1.77, and 1.68 eV by applying 1%, 2%, 3%, 4%, and 5% tensile strains, respectively. On the contrary, for compressive strains -1%, -2%, -3%, -4%, and -5%, the trend follows a blue-shift with the order, 2.40, 2.55, 2.69, 2.79, and 2.84 eV, respectively. This trend is in agreement with other theoretical studies, where the influence of strain on optical properties is investigated.[70, 71] Figure 4(b, e) presents the transmittance as a function of the biaxial strain for MWLH. The transimittance is almost 100% upto 1.5 eV, which then decreases afterwards. The trend of variation in its magnitude as a function of strain is illustrated. In Fig. 4(c), (f), the reflectance spectra as a function of strain are shown, indicating negligible reflectance of the incident light. Applying biaxial strain to the layered heterostructures is an effective approach to tuning the optoelectronic features of 2D laterally stitched monolayered heterostructures. The controlled modification of the optical properties via biaxial strain can offer tremendous applications in optoelectronic devices.[72-74]

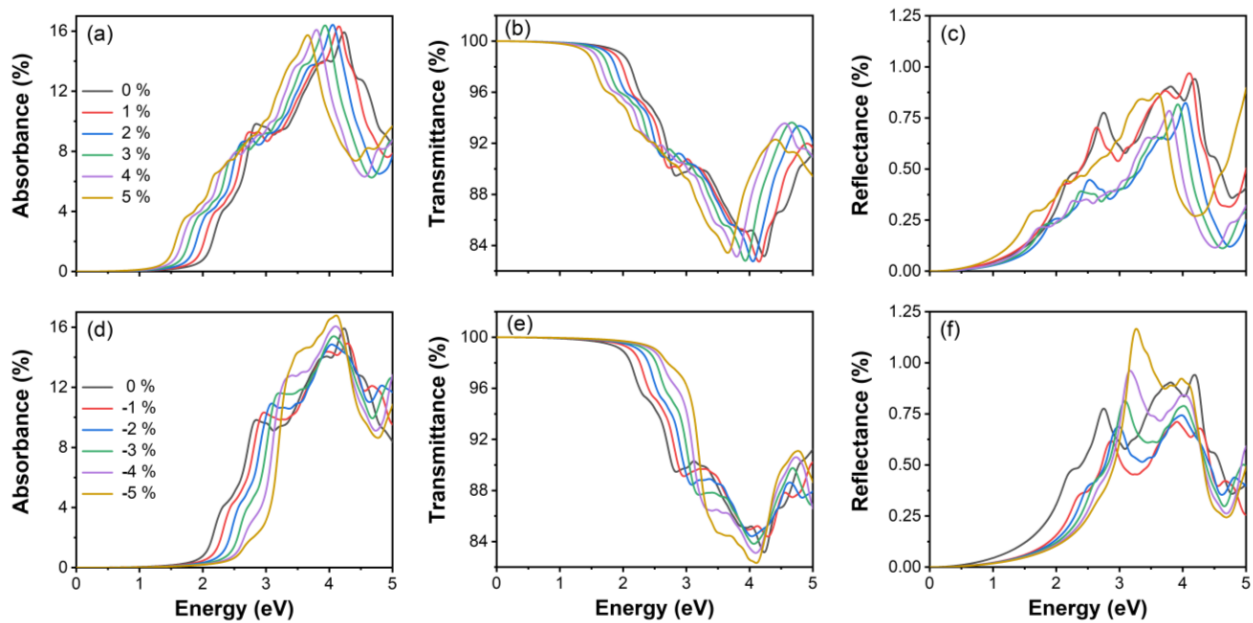

**Figure 4** The optical properties of MWLH as a function of bi-axial strain, the values of strains are shown in the insets of leftmost panel. (a, b, c) present the optical absorbance, transmittance and reflectance for the successive tensile strains, while (d, e, f) show the optical spectra when compressive strain is applied to the heterostructure.

## 4. Conclusions

In summary, based on the first-principles calculations, we investigated the structural stability, and electronic band structures of lateral $MoSi_2N_4/XSi_2N_4$ (X=W, Ti) 2D heterostructures. Specifically, the effect of biaxial strain on the electronic and optical properties of MWLH is addressed. Our results suggested that the $MoSi_2N_4/WSi_2N_4$ and $MoSi_2N_4/TiSi_2N_4$ lateral heterostructures are dynamically stable. It is concluded that the electronic and optical properties of MWLH can effectively be tuned using bi-axial strain engineering. The most significant outcome is that one can easily modulate the electronic band structures of MWLH, and induce transitions from indirect to direct bandgap semiconductor via moderate biaxial strain (less than ±5%). These modifications in the electronic band structures and optical spectra are associated with the changes in the bond lengths because of biaxial strain, principally that of Mo-N ($d_{W-N}$), W-N ($d_{Mo-N}$) and Si-N ($d_{Si-N}$). The designing of stable lateral heterostructures and the tuning of their electronic and optical attributes via strain engineering could be very promising in optoelectronics devices.

**Supplementary material**

See supplementary material for the band structures and the PDOS calculations.

**Note**

The authors declare no competing financial interest.

**Acknowledgments**

This work is supported by the Foundation for Polish Science through the international research agendas program co-financed by the European Union within the smart growth operational program. We acknowledge the access to the computing facilities of the Interdisciplinary Center of Modeling at the University of Warsaw, Grants No. G75-10, No. GB84-0, and No. GB84-7. The authors extend their appreciation to Riphah International University for funding this work under project number R-ORIC-21/FEAS-10.